\documentstyle[12pt]{article}
\font\goth=eufm10
\def\lb{\left[}
\def\rb{\right]}
\def\cc{{\hbox{C\hskip-0.2truecm I}}}
\def\nn{{\hbox{I\hskip-0.05truecmN}}}
\def\rr{{\hbox{I\hskip-0.2truecm R}}}
\def\zz{{\hbox{Z\hskip-0.2truecm Z}}}
\def\aa{{\cal A}}
\def\dd{{\cal D}}
\def\gg{\hbox{\goth g}}
\def\hh{{\cal H}}
\def\hhh{{\hbox{I\hskip-0.2truecm H}}}
\def\ff{{\cal F}}
\def\sss{{\cal S}}
\def\t{{\rm tr}}
\def\tt{{\rm tr}_\omega}
\def\ddd{\hbox{$\partial\!\!\!/$}}
\def\de{\hbox{\rm d}}
\def\dt{\hbox{$\delta_t$}}
\def\ul{\underline}

\def\bb{\begin{eqnarray}}
\def\ee{\end{eqnarray}}

\begin{document}

%\magnification=1200
%\catcode `@=11

\hsize 17truecm
\vsize 24truecm
\font\twelve=cmbx10 at 13pt
\font\eightrm=cmr8
\baselineskip 18pt
%\nopagenumbers
\begin{titlepage}
\centerline{\twelve CENTRE DE PHYSIQUE THEORIQUE}
\centerline{\twelve CNRS - Luminy, Case 907}
\centerline{\twelve 13288 Marseille Cedex}
%\null
\vskip 8truecm

\centerline{\twelve CONNES' MODEL BUILDING KIT}

\bigskip

\begin{center}
{\bf Thomas SCH\"UCKER}\footnote{and Universit\'e de
Provence}\\
{\bf Jean-Marc ZYLINSKI}\footnote{and
Universit\'e d'Aix-Marseille II}
\end{center}

\vskip 2truecm
\leftskip=1cm
\rightskip=1cm
\centerline{\bf Abstract}

\medskip

Alain Connes'  applications of non-commutative
geometry to interaction physics are described for the
purpose of model building.

\vskip 2truecm

\vskip 3truecm

\noindent October 1993

\noindent CPT-93/P.2960

\end{titlepage}

The possibilities of the Yang-Mills-Higgs model
building kit have been explored thoroughly in the last
30 years and the  standard $ SU(3) \times SU(2) \times
U(1) $ model of strong and electroweak  interactions
has emerged quite uniquely as adequate description of
high energy phenomena. On the mathematical side
this kit relies on  two  ingredients: the differential
forms on spacetime together with their  structure of
differential algebra and a Lie group represented on a
finite dimensional internal space. These two
ingredients are thrown  together in a tensor product
to yield a gauge invariant action. In  Connes' approach
[1] both spacetime and internal space are described
by involution algebras and their tensor product
produces a special class  of Yang-Mills actions with
spontaneous symmetry breaking. In this  class the
fermionic mass matrix is naturally unified with the
Dirac  operator, the Higgs scalars with the gauge
bosons and the Higgs  potential with the Yang-Mills
Lagrangian. While the introduction of  Lie groups is ad
hoc, involution algebras have a profound
mathematical motivation in this context,
non-commutative geometry.  A model builder however
who is willing to accept Connes' rules  can very well do
with a minimum of mathematics to be introduced
below.

Let us quickly review input and
output of the Yang-Mills-Higgs kit.  To get started we
have to commit ourselves to the following choices:
\begin{itemize}
\item
 a (finite dimensional) real, compact Lie group $G$,
\item a positive definite, bilinear
invariant form on the Lie algebra  $\gg$ of $G$. This
choice is parametrized by a few positive numbers
$g_i$, the coupling constants.
\item a
(faithful, unitary) representation $\hh_L$ for left
handed fermions  (spin$1\over2$),
\item
a representation $\hh_R$ for right handed fermions,
\item
 a representation $\hh_S$ for
scalars (spin 0),
\item
 an invariant
positive polynomial of order 4 on the representation
space of the scalars, this polynomial is denoted by
$V(\phi),  \phi \in \hh_S$, the Higgs potential,
\item
 one complex number or Yukawa
coupling $g_Y$ for every singlet in the decomposition
of the representation
\bb  \left(\hh_L^{\ast}\otimes
\hh_R\otimes \hh_S\right) \oplus\left(
     \hh_R^\ast\otimes \hh_L\otimes \hh_S\right). \ee
\end{itemize}
With these ingredients the
popular Yang-Mills-Higgs algorithm produces the
following output:
\begin{itemize}
 \item the particle
spectrum, roughly one particle for each basis vector in
\bb \gg^\cc\oplus \hh_L\oplus \hh_R\oplus \hh_S. \ee
The basis elements of the complexified Lie algebra
$\gg^\cc$  are the gauge bosons, they have spin 1.
\item
the particle masses,
\item
 the interactions, e.g. charges,
Kobayashi-Maskawa matrix...
\end{itemize}
Trial
and error, that is guessing an input, calculating the
output and comparing with experiment, has singled
out the standard model of electroweak and strong
interactions:
\bb G =  SU(3) \times SU(2) \times U(1) \ee
with three coupling constants $ g_3, g_2, g_1, $
\bb
\hh_L = \lb (1,2,-1)\oplus (3,2,{1 \over 3}) \rb \
{\times} \  3 \ee
\bb \hh_R = \lb (1,1,-2)\oplus
(3,1,{4\over 3})\oplus (3,1,-{2\over 3}) \rb \  {\times}
\  3 \ee
\bb \hh_S = (1,2,-1) \ee
where $(n_3,n_2,y)$
denotes a tensor product of an $n_3$ dimensional
representation under $SU(3)$, an $n_2$ dimensional
representation under $SU(2)$ and the one dimensional
representation of $U(1)$ with hypercharge $y$:
\bb
\rho(e^{i\theta}) = e^{i3y\theta}, \qquad 3y\in\zz,\
\theta \in [0,2\pi)\ee
\bb V(\phi) = \lambda (\phi^\ast
\phi)^2 - {{\mu ^2 }\over 2}\phi^\ast \phi,
\qquad \phi \in
\hh_S \quad \lambda, \mu > 0.\ee
There are 27 Yukawa
couplings in the input, enough to allow for an
arbitrary fermionic mass matrix (fermion masses and
Kobayashi-Maskawa matrix) and an arbitrary scalar
mass in the  output.

 In a first step
we only consider a particular subset of Connes' model
building kit, where spacetime and internal space
come in a tensor product. This subset compares
naturally with the Yang-Mills-Higgs model building
kit. As before the input concerns only the (finite
dimensional) internal space:
\begin{itemize}
\item
 an
associative involution algebra $\aa$ with unit 1,
\item
 two representations $\hh_L$ and
$\hh_R$ of $\aa$, \item
 a mass matrix
$M$ i.e. a linear map
$ M: \hh_R \longrightarrow
\hh_L, $
\item
a certain number of
coupling constants depending on the degree of
reducibility of $ \hh_L\oplus \hh_R. $
\end{itemize}

The data $(\hh_L,\hh_R,M)$ plays
a fundamental role in non-commutative geometry
where it generalizes the Dirac operator. It is called
K-cycle.

 On the output side we
find a complete action of Yang-Mills-Higgs type with
$G$ the group of unitary elements in $\aa$:
\bb G =
\{g\in \aa,\quad gg^\ast =g^*g=1\} \ee
or possibly a
subgroup thereof. In other words in Connes' approach
the representation space $\hh_S$ of scalars sits at the
output end  together with the representation
space $ \gg^\cc$ of the gauge bosons. Likewise the
quartic Higgs potential together with the
Klein-Gordon Lagrangian for the scalars are found in
the output accompanying the quartic Lagrangian for
the gauge bosons.

In the following
we first discuss Connes' rules for a finite  dimensional
(internal) space. Thereby we avoid the difficulties of
functional analysis, only prerequisites being linear
algebra, e.g. matrix  multiplication. The second
chapter deals with a particular infinite dimensional
algebra, the functions on spacetime, and the Dirac
operator. Model builders are well acquainted with this
part of  mathematics which also motivates Connes'
rules. The third chapter is simply the tensor product
of the first two. As we go along, the general rules are
illustrated by an example. The simplest typical  one we
know strikingly  resembles the
Glashow-Salam-Weinberg model.

\section{The internal space}

An involution algebra $\aa$ is an
associative algebra with unit 1 and an involution $^*$
i.e. an antilinear map from $\aa$ into  itself,
\bb
(a+b)^* = a^*+b^*, \qquad a, b \in \aa,\ee
\bb (\lambda a)^*=\lambda ^* a^*, \qquad
\lambda \in \rr \  {\rm
or} \ \cc,  \ee
with the properties:
\bb a^{**} = a,\ee
\bb1^*= 1,\ee
\bb (ab)^*=b^*a^*.\ee
The two classical examples
of involution algebra are:
\hfil\break\noindent
1.
$M_n(\cc)$, the algebra of $n\times n$ matrices with
complex entries. The involution is transposition and
complex conjugation.
\hfil\break\noindent
2. The
algebra of functions from a manifold (spacetime) into
the complex numbers. The involution is just complex
conjugation.
\hfil\break\noindent
The second
example is infinite dimensional and commutative,
while the first one is finite dimensional and non-%
commutative.  Note that in this context the word non-%
abelian is non-fashionable.

 A
representation $\rho$ of $\aa$ over a Hilbert space
$\hh$ is a  homomorphism from $\aa$ into the
operators on $\hh$:
\bb \rho : \aa
\longrightarrow&{\rm End}(\hh)\cr
           a \longmapsto& \rho(a) .\ee
Here
homomorphism refers to all given structures, addition,
scalar multiplication, multiplication, unit and
involution:
\bb    \rho (a+b)=\rho (a)+\rho (b), \ee
 \bb
\rho (\lambda a) = \lambda \rho(a), \label{rho} \ee
\bb    \rho
(ab)=\rho (a) \rho (b), \ee
\bb    \rho (1)=1, \ee
\bb    \rho
(a^*)=\rho (a)^*.\ee
All representations will be
supposed faithful, i.e. injective.

Model builders should note that the two choices,
algebra and  representation, are much more restricted
than the choice of a group and a representation. Not
every group $G$ is the group of unitary elements in
an algebra $\aa$. The phenomenologically important
$SU(3)$ is an example to which we shall have to come
back. Every algebra representation yields a group
representation of its group of unitaries, but most of
these group representations cannot be obtained from
an algebra representation. For example the group
$SU(2)$ has unitary representations of any dimension,
1, 2, 3... while the algebra of quaternions whose group
of unitaries is $SU(2)$ admits only one irreducible
representation, the one of dimension 2. Indeed the
tensor product of two algebra representations is not a
representation because  compatibility with the linear
structure is lost. Also the popular singlet
representation, ``dark matter'', is not available.

A K-cycle $(\hh, \dd, \chi)$ of an
algebra $\aa$ consists of a (faithful) representation
$\rho$ on a Hilbert space $\hh$, of a  self adjoint
operator $\chi$ on $\hh$, called chirality, satisfying
\bb    \chi^2 = 1 \ee
and of a self adjoint operator $\dd$ on
$\hh$, the generalized Dirac operator. Furthermore
we suppose that $\rho (a) $ is even:
\bb    \rho(a) \chi =
\chi \rho(a) \ee
 for all $a \in \aa$ and that $\dd $ is
odd:
\bb    \dd \chi = -\chi \dd .\ee
 In the infinite
dimensional case there will be additional conditions to
be satisfied by $\dd$.

In other
words the representation $\rho$ is reducible and
decomposes into a left handed and a right handed part
$\rho_L$ and $\rho_R$  living on the left handed and
right handed Hilbert spaces
\bb   \hh_L :=
{{1-\chi}\over2} \hh,\ee
\bb   \hh_R := {{1+\chi}\over2}
\hh.\ee
For a finite dimensional Hilbert space we can
pick a basis such that
\bb    \chi = \pmatrix {1_L&0\cr
0&-1_R}.\ee
Then
\bb   \rho = \pmatrix {\rho_L & 0 \cr 0&
\rho_R},\ee
 \bb   {\dd}= \pmatrix {0 & M \cr
 M^\ast & 0},\ee
with $M$ a matrix of size
$\rm{dim}\hh_L \times \rm{dim}\hh_R$, the mass
matrix.

{\bf Example:} Let $\aa :=
M_2(\cc)\oplus \cc$ and denote its elements by
$(a,b),$ $a$ being a $2\times 2$ matrix and $b$ a
complex number. We define a K-cycle by
\bb    \hh_L :=
\cc^2, \qquad \hh_R := \cc^2,\ee
\bb    \rho_L(a,b) := a,
\qquad \rho_R(a,b) := B\ee
with
\bb   B :=  \pmatrix {b&0\cr
0&\bar b}\ee
 and
 \bb   M := \pmatrix {m_1&0\cr 0&m_2},
\quad m_1,m_2 \in \cc.\ee
We shall always assume
$|m_1|\ne |m_2|$. Later on it will be convenient to
have $M$ of  maximal rank. Should one of the $m_i$
vanish this can still be achieved by the following trick
sometimes called leptonic  reduction: Suppose $m_2=0$.
We simply omit the second dimension  of $\hh_R$
which becomes one dimensional. Then $\rho(a,b)$
and  $\dd$ are $3\times3$ matrices and
\bb   M = \pmatrix
{m_1\cr 0} \ee
is reduced to a column vector of maximal
rank.

Note the appearance of the
complex conjugation in $\bar b$ which by equation
(\ref{rho}) means that we
must consider the algebra as defined over the real
numbers.

Given an (involution)
algebra and a K-cycle we now want to  construct a
differential algebra. Let us recall the axioms of a
differential algebra $\Omega$. It is a graded vector
space
\bb   \Omega =\bigoplus_{p\in\nn_0} \Omega^p.\ee
We denote its associative product by juxtaposition:
\bb
\Omega^p \times \Omega^q \longrightarrow&
\Omega^{p+q}\cr
           (\phi,\psi) \longmapsto& \phi\psi .\ee
Furthermore $\Omega$ is equipped with a differential
$\delta$  that is a linear map
\bb
\delta:\quad\Omega^p \longrightarrow&
\Omega^{p+1}\cr
           \phi \longmapsto& \delta\phi  \ee
 with two
properties. It is nilpotent,
\bb   \delta^2=0\ee
 and obeys a
graded Leibniz rule,
\bb   \delta
(\phi\psi)=(\delta\phi)\psi+(-1)^p\phi\delta\psi,
\quad \phi\in\Omega^p.\ee

The differential algebra is called (graded)
commutative if in addition its product satisfies
\bb   \phi\psi=(-1)^{pq}\psi\phi,\quad
\phi\in\Omega^p,\ \psi\in\Omega^q.\ee

{}From our given algebra $\aa$ we now construct first
an auxiliary differential algebra $\hat{\Omega} \aa,$
the so called universal differential envelop of $\aa$:
\bb   \hat{\Omega}^0\aa := \aa,\ee
$\hat{\Omega}^1\aa$ is
generated by symbols $\delta a,\quad a \in \aa$ with
relations
\bb   \delta 1 = 0 \ee
\bb    \delta(ab) = (\delta
a)b+a\delta b.\ee
 Therefore $\hat\Omega^1\aa$ consists
of finite sums of terms of the form $a_0\delta a_1,$
\bb   \hat\Omega^1\aa = \left\{ \sum_j a^j_0\delta
a^j_1,\quad a^j_0, a^j_1\in \aa\right\}\ee
and likewise
for higher $p$
\bb   \hat\Omega^p\aa = \left\{ \sum_j
a^j_0\delta a^j_1...\delta a^j_p ,\quad a^j_q\in
\aa\right\}.\ee
 The differential $\delta$ is defined by
\bb   \delta(a_0\delta a_1...\delta a_p) :=
   \delta a_0\delta a_1...\delta a_p.\ee

Two remarks: The universal differential envelop
$\hat\Omega\aa$ of a commutative algebra $\aa$ is
not necessarily graded commutative. The
universal differential envelop of any algebra has no
cohomology. This means that every closed form
$\hat\phi$ of degree  $p\geq 1,\quad
\delta\hat\phi=0,$ is exact, $\hat\phi = \delta\hat\psi$
for some $(p-1)$form $\hat\psi.$

The involution $^*$ can be
extended from the algebra $\aa$ to its universal
differential envelop $\hat\Omega^1\aa$ by putting
\bb   (\delta a)^* := \delta(a^*) =:\delta a^*\ee
and of course
\bb (\phi\psi)^*=\psi^*\phi^*.\ee
Note that
Connes has a minus sign in the above relation which
amounts to replacing $\delta$ by $i\delta.$

Our next step is to extend the
representation $\rho$ from the  algebra $\aa$   to its
universal differential envelop $\hat\Omega\aa$. This
extension is the central piece of Connes' algorithm
and deserves a new name:
\bb  \pi :
\hat\Omega\aa  \longrightarrow&{\rm End}(\hh)\cr
           \hat\phi \longmapsto& \pi(\hat\phi) \ee
$$\pi(a_0\delta a_1...\delta a_p) :=
(-i)^p\rho(a_0)[\dd,\rho(a_1)] ...[\dd,\rho(a_p)].$$
Note that in Connes' notations there is no factor
$(-i)^p$ on the rhs. A straight-forward calculation
shows that $\pi$ is in fact a  representation of
$\hat\Omega\aa$ as involution algebra, but the
differential is not respected unless the generalized
Dirac operator $\dd$ satisfies
\bb   [\dd,[\dd,\cdot]]\equiv
0.\ee
 Otherwise there can be forms $\hat\phi$ with
$\pi(\hat\phi)=0$ and  $\pi(\delta\hat\phi)\not= 0$.
By dividing out these unpleasant forms we shall
construct a new differential algebra $\Omega\aa$, the
real thing. It will of course depend on the chosen
K-cycle.

The kernel of $\pi$ is a
(bilateral) ideal in the image $\pi(\hat \Omega\aa)$.
We turn it into a differential ideal $J$ ($J$ for junk):
\bb   J := \ker\pi + \delta\ker\pi =: \bigoplus_p J^p\ee
with
\bb   J^p = (\ker\pi)^p + \delta(\ker\pi)^{p-1}\ee
 and
divide it out:
\bb   \Omega\aa := {{\hat\Omega\aa}\over
J}.\ee
The differential $\delta$ on $\hat\Omega\aa$
passes to the quotient where it is again denoted by
$\delta$. Degree by degree we have:
\bb   \Omega^0\aa =
\hat\Omega^0\aa \cong \rho(\aa)\ee
because $\rho$ is
faithful and $J^0=(\ker\pi)^0=0$,
\bb   \Omega^1\aa =
{{\hat\Omega^1\aa}\over{(\ker\pi)^1}} \cong
\pi(\hat\Omega^1\aa)\ee
because $J^1=(\ker\pi)^1$,
and in degree $p\geq2$
\bb   \Omega^p\aa \cong
{{\pi(\hat\Omega^p\aa)}\over
{\pi(\delta(\ker\pi)^{p-1})}}.\ee
While $\hat\Omega\aa$ has no cohomology
$\Omega\aa$ does in general. In fact let us anticipate,
if $\ff$ is the algebra of complex functions on a
compact spin manifold $M$ of even dimensions and if
the K-cycle is obtained from the Dirac operator then
$\Omega\ff$ is de Rham's differential algebra of
differential forms on $M$.

We come back to our finite
dimensional case. Remember that the elements of the
auxiliary differential algebra $\hat\Omega\aa$ that
we introduced for book keeping purposes only are
abstract entities defined in terms of symbols and
relations. On the other hand, with the above
isomorphisms, the elements of $\Omega\aa$, the
``forms'', are operators on the Hilbert space $\hh$ of
the K-cycle, i.e. concrete matrices of complex
numbers.

{\bf Examples:} Before
continuing our example above let us mention  a class
of trivial examples that deserve the  name {\it vector
like models}. The algebra $\aa$ is arbitrary, left and
right representations are equal $\rho_L = \rho_R$,
and the mass  matrix appearing in the ``Dirac'' operator
$\dd$ is a multiple of the  identity matrix $M = \lambda
1, \  \lambda\in\cc$. We shall see  that these models
will produce Yang-Mills theories with unbroken
parity and unbroken gauge symmetry as
electromagnetism and chromodynamics. Any vector
like model has trivial differential algebra,
\bb
\Omega^0\aa = \aa,\ee
 \bb   \Omega^p\aa = 0, \quad
p=1,2,3,...\ee
 Coming back to our serious example recall:
\bb    \rho(a,b) = \pmatrix {a&0\cr 0&B}, \qquad  \dd =
\pmatrix {0&M\cr M^*&0}.\ee
We need the commutator
\bb [\dd,\rho(a,b)] &= \pmatrix
{0&M\rho_R(b)-\rho_L(a) M\cr M^*\rho_L(a)-
\rho_R(b)
M^*&0}\cr  &=\pmatrix {0&MB-aM\cr
M^*a-BM^*&0}\cr &= \pmatrix {0&-(a-B)M\cr
M^*(a-B)&0}\ee
where in the last equality we have
used that $M$ and $B$ are  diagonal and commute. A
general element in $\Omega^1\aa$ is of  the form
\bb
\pi((a_0,b_0)\delta(a_1,b_1)) = -i\pmatrix
{0&-a_0(a_1-B_1)M\cr M^*B_0(a_1-B_1)&0}\ee
 and as
vector space
\bb   \Omega^1\aa = \left\{i\pmatrix
{0&hM\cr M^*\tilde h^*&0},\ h,\tilde h\in
M_2(\cc)\right\}.\ee
  Likewise a general element in
$\pi(\hat\Omega^2\aa)$ is
\bb
&&\pi((a_0,b_0)\delta(a_1,b_1)\delta(a_2,b_2))=\ee
$$\pmatrix {a_0(a_1-B_1)MM^*(a_2-B_2)&0\cr
0&M^*B_0(a_1-B_1)(a_2-B_2)M}=$$
$$\pmatrix {\Sigma a_0(a_1-B_1)(a_2-B_2)+ \Delta
a_0(a_1-B_1)\sigma_3(a_2-B_2)&0\cr
0&M^*B_0(a_1-B_1)(a_2-B_2)M}$$
 where we have
used the decomposition
\bb   MM^* = \pmatrix
{|m_1|^2&0\cr 0&|m_2|^2} = \Sigma 1+\Delta\sigma_3\ee
with
\bb   \Sigma := {1\over2}(|m_1|^2+|m_2|^2),\ee
\bb   \Delta := {1\over2}(|m_1|^2-|m_2|^2),\ee
\bb   \sigma_3 :=
\pmatrix {1&0\cr 0&-1}.\ee
A general element in
$(\ker\pi)^1$ is a finite sum of the form
\bb    \sum_j
(a^j_0,b^j_0)\delta (a^j_1,b^j_1) \ee
 with the two
conditions
\bb   \left[ \sum_j a^j_0(a^j_1-B^j_1)\right]M =
0,\ee
\bb   M^*\left[ \sum_j B^j_0(a^j_1-B^j_1)\right] = 0.\ee
Therefore the corresponding general element in
$\pi(\delta (\ker\pi)^1)$ is
\bb   \pmatrix
{\Sigma\sum_j(a^j_0-B^j_0)(a^j_1-B^j_1)+
\Delta\sum_j(a^j_0-B^j_0)\sigma_3(a^j_1-B^j_1)&0\cr
0&0} \ee
 still subject to the two conditions.  Recall that
$\Delta \ne 0$ by assumption and we have  the
following inclusion
\bb   \pi(\delta(\ker\pi)^1) \supset &
\left\{\pmatrix {\Delta\sum_j a^j_0\sigma_3
a^j_1&0\cr 0&0},\quad \sum_j a^j_0a^j_1=0 \right\}\cr
 &= \left\{ \pmatrix {\Delta k&0\cr 0&0}, \quad k \in
M_2(\cc)\right\}.\ee
   To prove the last equality we
note that the subspace is a bilateral  ideal in the rhs.
Furthermore the subspace contains  the non-zero
element with:
\bb   a_0 := \pmatrix {0&0\cr 1&-1},\ee
\bb   a_1
:= \pmatrix {1&1\cr 1&1},\ee
 \bb   a_0a_1 = 0,\ee
\bb   a_0\sigma_3 a_1 = \pmatrix {0&0\cr 2&2}.\ee
The
algebra $M_2(\cc)$ being simple the subspace is the
entire algebra. Consequently
\bb   \pi(\delta(\ker\pi)^1)
= \left\{ \pmatrix {\Delta k&0\cr 0&0}, \quad k \in
M_2(\cc)\right\}.\ee
 Now  we have to compute the
quotient
\bb   \Omega^2\aa =
{{\pi(\hat\Omega^2\aa)}\over
{\pi(\delta(\ker\pi)^{1})}}.\ee
We are tempted to conclude
\bb   \Omega^2\aa = \left\{
\pmatrix {0&0\cr 0&M^*cM}, \quad c \in
M_2(\cc)\right\}.\ee
The problem with this conclusion
is that a quotient of vector spaces  consists of classes
and is not canonically a subspace. The situation is
simpler if  our vector space comes equipped with a
scalar product, in which  case there is a privileged
representative in each class, the one
orthogonal to the subspace $\pi(\delta(\ker\pi)^{1})$.
This  canonical choice allows to consider the quotient
as subspace.

Since the elements of
$\pi(\hat\Omega\aa)$ are operators on the Hilbert
space $\hh$, i.e. concrete matrices, they have a
natural scalar product defined by
\bb   <\hat\phi,\hat\psi> := \t (\hat\phi^*\hat\psi),
\quad  \hat\phi, \hat\psi \in \pi(\hat\Omega^p\aa)\ee
for forms of equal degree and zero for the scalar
product of two  forms of different degree. With this
scalar product  $\Omega\aa$ is a subspace of
$\pi(\hat\Omega\aa)$ and by  definition orthogonal to
$J = \ker\pi + \delta\ker\pi$. Now our  conclusion in
the example makes sense. As a subspace $\Omega\aa$
inherits a scalar product which deserves a special
name ( , ). It is  given by
\bb   (\phi,\psi) =
\t(\phi^*P\psi), \quad \phi, \psi \in \Omega^p\aa\ee
where $P$ is the orthogonal projector in
$\pi(\hat\Omega\aa)$  onto the ortho\--complement
of $J$ and $\phi$ and $\psi$ are any  representatives
in their classes. Again the scalar product vanishes  for
forms with different degree.

Let
us recapitulate our example:
\bb   \Omega^0\aa = \left\{
\pmatrix {a&0\cr 0&B}, \quad a \in M_2(\cc),\
B=\pmatrix {b&0\cr 0&\bar b}\right\},\ee
\bb   \Omega^1\aa = \left\{i\pmatrix {0&hM\cr M^*\tilde
h^*&0},\  h,\tilde h\in M_2(\cc)\right\},\ee
\bb   \Omega^2\aa = \left\{ \pmatrix {0&0\cr 0&M^*cM},
\quad c \in M_2(\cc)\right\}.\ee
Since $\pi$ is a
homomorphism of involution algebras the product  in
$\Omega\aa$ is given by matrix multiplication
followed by the  projection
\bb P=\pmatrix{0&0\cr
0&1}\ee
 and the involution  is
given by transposition complex conjugation. It is in
order to
calculate the  differential $\delta$ that we need the
complicated construction  above:
\bb \delta
:\Omega^0\aa&\longrightarrow&\Omega^1\aa\cr\cr
 \pmatrix
{a&0\cr 0&B} &\longmapsto& i\pmatrix {0&(a-B)M\cr
-M^*(a-B)&0},\\ \cr\cr
\delta :
\Omega^1\aa&\longrightarrow&\Omega^2\aa\cr\cr
 i\pmatrix {0&hM\cr M^*\tilde
h^*&0} &\longmapsto& \pmatrix {0&0\cr
0&M^*(h+\tilde h^*)M}.\ee
Let us note that
not every example can be calculated as explicitly  as
the above one. Just add a non-diagonal entry in the
mass matrix  $M$ of our example and $\delta$ will not
have the simple form  indicated.

We are now ready to make a first contact with gauge
theories.  Consider the vector space of antihermitian
1-forms
\bb   \left\{ H\in \Omega^1\aa, H^*=-H
\right\}.\ee
A general element $H$ is of the form
\bb   H =
i\pmatrix {0&hM\cr M^*h^*&0}, \quad
h:\hh_L\rightarrow\hh_R.\ee
 These elements are
called Higgses or gauge potentials.  In fact the space of
gauge potentials carries an affine  representation of
the group of unitaries
\bb    G = \{g\in \aa, gg^\ast
=g^*g=1\} \ee
defined by
\bb H^g &:=&\
\rho(g)H\rho(g^{-1})+\rho(g)\delta \rho(g^{-1}) \cr
       &=&\ \rho(g)H\rho(g^{-1})+(-i)\rho(g)[\dd,\rho
(g^{-1})] \cr
       &=:& i\pmatrix {0&h^gM\cr M^*(h^g)^*&0}\ee
with
\bb    h^g = \rho_L(g)[h-1]\rho_R(g^{-1}) + 1.\ee
 $H^g$ is
the ``gauge transformed of $H$''.  As usual every gauge
potential $H$ defines a covariant derivative  $\delta
+H$, covariant under the left action of $G$ on
$\Omega\aa$:
\bb   ^g\psi := \rho(g)\psi, \quad
\psi\in\Omega\aa\ee
 which means
\bb   (\delta+H^g)\
^g\psi = \ ^g\lb(\delta+H)\psi\rb.\ee
 As usual we define the
curvature $C$ of $H$ by
\bb   C := \delta H+H^2\ \in
\Omega^2\aa.\ee
Note that here and later $H^2$ is considered as element
of $\Omega^2\aa$ which means it is the projection $P$
applied to $H^2\in \pi(\hat\Omega^2\aa)$.
The curvature $C$ is a hermitian 2-form with {\it
homogeneous} gauge transformations
 \bb   C^g :=
\delta(H^g)+(H^g)^2 = \rho(g) C \rho(g^{-1}).\ee
Finally we define the ``Higgs potential'' $V(H)$, a
functional on the  space of gauge potentials, by
\bb   V(H)
:= (C,C) = \t[(\delta H+H^2)P(\delta H+H^2)].\ee
It is a
polynomial of degree 4 in $H$ with real, non-negative
values.  Furthermore it is gauge invariant,
\bb    V(H^g)
= V(H)\ee
 because of the homogeneous transformation
property of the  curvature $C$ and because the
orthogonal projector $P$ commutes  with all gauge
transformations
\bb   \rho(g)P =P\rho(g).\ee
The
transformation law for $H$ motivates the following
change of  variables
\bb   \Phi := i\pmatrix {0&\phi M\cr
M^*\phi^*&0} := H-i\dd.\ee
In other words
 \bb   \phi =h-1.\ee
The new variable $\Phi$ transforms
homogeneously
\bb   \Phi^g = \rho(g)\Phi\rho(g^{-1})\ee
or
\bb   \phi^g = \rho_L(g)\phi\rho_R(g^{-1})\ee
 where
the differential is of course considered gauge
invariant
\bb   \dd^g = \dd.\ee

In our
example the Higgs $h$ is a complex $2\times2$ matrix,
which  in terms of $\phi$ decomposes under gauge
transformations into  two complex doublets, the two
column vectors $\phi_1$ and  $\phi_2$ of $\phi$. The
curvature is readily calculated
\bb   C := \delta H+H^2 =
\pmatrix {0&0\cr 0&M^*cM}\ee
 with
\bb   c = h+h^*-h^*h
= 1-\phi^*\phi.\ee
The Higgs potential is
\bb V(H)&=& \t\lb C^2\rb =
\  \t\lb(
M^*(1-\phi^*\phi)M\rb^2 \cr
            &=&|m_1|^4+|m_2|^4+|m_1|^4(\phi_1^*\phi_1)^2+
                                           |m_2|^4(\phi_2^*\phi_2)^2 \cr
             &&\ -2|m_1|^4\phi_1^*\phi_1
                -2|m_2|^4\phi_2^*\phi_2
+2|m_1|^2|m_2|^2(\phi_1^*\phi_2)(\phi_2^*\phi_1).\ee
Its most interesting feature is that it breaks the gauge
symmetry  spontaneously. Indeed the only gauge
invariant point in the space of  gauge potentials or
Higgses is $\phi=0$. This point is not a  minimum of
the Higgs potential.

\section{Spacetime}

In this chapter our algebra is
infinite dimensional, the algebra of  differentiable,
complex valued functions on spacetime $M$
\bb   \ff :=
{\cal C}^\infty(M).\ee
 The K-cycle is defined by the
Dirac operator. We sketch how the  differential
algebra $\Omega\ff$ reproduces the ordinary
differential forms on $M$. For simplicity spacetime is
taken flat, compact and Euclidean. To  define the  Dirac
operator we also need a spin structure. We denote by
$\sss$  the Hilbert space of the K-cycle. $\sss$ consists
of square integrable  spinors
\bb   \Psi = \pmatrix{
\Psi_1(x) \cr \Psi_2(x) \cr \Psi_3(x) \cr \Psi_4(x) }.\ee
The representation of $\ff$ on $\sss$ is simply by
multiplication  and is denoted by ${\ul\cdot}$ ,
\bb   ({\ul
f}\Psi)(x) := f(x)\Psi(x), \quad f\in \ff,\ \Psi\in\sss.\ee
The odd operator of the K-cycle is the (true) Dirac
operator
\bb   \ddd\Psi := i\sum_{\mu=0}^3\gamma^\mu
{\partial\over{\partial x^\mu}}\Psi.\ee
 Our gamma
matrices are self adjoint,
\bb   \gamma ^0 :=  \pmatrix {1
& 0 & 0 & 0 \cr
 0 & 1 & 0 & 0 \cr
 0 & 0 & -1 & 0 \cr
 0 & 0 & 0 & -1} \qquad \gamma ^1 :=  \pmatrix {0 & 0 &
0 & i \cr
 0 & 0 & i & 0 \cr
 0 & -i & 0 & 0 \cr
 -i & 0 & 0 & 0}\ee
 \bb   \gamma ^2 :=  \pmatrix {0 & 0 & 0 &
1 \cr
 0 & 0 & -1 & 0 \cr
 0 & -1 & 0 & 0 \cr
 1 & 0 & 0 & 0} \qquad \gamma ^3 :=  \pmatrix {0 & 0 & i
& 0 \cr
 0 & 0 & 0 & -i \cr
 -i & 0 & 0 & 0 \cr
 0 & i & 0 & 0}.\ee
They satisfy the anticommutation
relation
\bb   \gamma ^\mu \gamma ^\nu +\gamma ^\nu
\gamma ^\mu =  2\eta ^{\mu \nu }1\ee
 with the flat
Euclidean metric
\bb   \eta = \pmatrix {1 & 0 & 0 & 0 \cr
 0 & 1 & 0 & 0 \cr
 0 & 0 & 1 & 0 \cr
 0 & 0 & 0 &1 }.\ee
 The chirality operator is by definition
\bb    \gamma_5 :=
\gamma^0\gamma^1\gamma^2\gamma^3 =
 \pmatrix {0 & 0 & -1 & 0 \cr
 0 & 0 & 0 & -1 \cr
 -1 & 0 & 0 & 0 \cr
 0 & -1 & 0 & 0}.\ee
It is self adjoint, its square is one as
postulated and since it anticommutes with all other
gamma matrices
\bb   \gamma^\mu\gamma_5+\gamma_5\gamma^\mu =
0,\ee
the Dirac operator is odd
\bb   \ddd\gamma_5+\gamma_5\ddd = 0.\ee
 The Dirac
operator $\ddd$ has additional algebraic properties:
\begin{itemize}
\item
 The commutator $[\ddd,{\ul f}]$ is
a bounded operator for all  $f\in\ff$.
\item
The spectrum of $\ddd$ is discrete
and the eigenvalues  $\lambda_n$ of $|\ddd|$ arranged
in an increasing sequence are of  order $n^{(1/d)}$
for $d$ dimensional manifolds $M$, $d$ even,  for us
$d=4$.
\end{itemize}
 These two properties are
added to the axioms of an abstract  K-cycle. They will
be needed later to define a trace.

We denote by $\de$ the differential in the universal
differential  envelop $\hat\Omega\ff$ and by $\pi_D$,
$D$ for Dirac, the algebra  homomorphism
\bb
\pi_D: \hat\Omega\ff  \longrightarrow&{\rm
End}(\sss)\cr
           \hat\phi \longmapsto& \pi_D(\hat\phi) .\ee
\bb   \pi_D(f_0\de f_1...\de f_p) := (-i)^p{\ul
f_0}[\ddd,{\ul f_1}] ...[\ddd,{\ul f_p}].\ee
We need the
commutator
\bb [\ddd,{\ul f}]\Psi &=&
i\gamma^\mu{\partial\over{\partial x^\mu}} (f\Psi)-
if\gamma^\mu{\partial\over{\partial x^\mu}}\Psi
\cr  &=& i\lb\gamma^\mu{\partial\over{\partial
x^\mu}}f\rb\Psi.\ee
(We use Einstein's summation
convention.) Therefore
\bb   [\ddd,{\ul f}] =
i\gamma^\mu{\partial\over{\partial x^\mu}}f  =:
i\gamma(\de f)\ee
with
\bb   \de f =
\lb{\partial\over{\partial x^\mu}}f\rb\de x^\mu.\ee
At
this point already we see that the restriction to flat
spacetime  can be dropped. The Dirac operator on
curved manifolds
\bb   i\gamma^\mu(x)\lb{\partial\over{\partial
x^\mu}}+ \omega_\mu\rb\ee
 differs from the flat one in two
respects, the gamma matrices are $x$  dependent, no
problem in the above commutator, and an additional
algebraic term, a spin connection $\omega =
\omega_\mu\de x^\mu$ valued in $so(4)$ appears but
drops out  from the commutator. Since the Dirac
operator only shows up in  commutators Connes'
algorithm works on any Riemannian manifold.

The representation of functions
by multiplication on spinors is  faithful, of course, and
\bb   \Omega^1\ff \cong \pi_D(\hat\Omega^1\ff)\ee
A
general element of the rhs is a finite sum of terms
\bb
\pi_D(f_0\de f_1),\quad f_0,f_1\in\ff.\ee
It is identified
with the differential 1-form on $M$
\bb     f_0\de f_1
\quad \in\Omega^1M.\ee
For 2-forms the situation is
less trivial, we must compute the junk
$\pi_D(\de(\ker\pi_D)^1)$. Consider
\bb   h^{-1}\de
h+h\de h^{-1}\ee
an element in $\hat\Omega^1\ff$ where
$h\in\ff$ is a non-vanishing  function,
$h^{-1}(x)=1/h(x)$. As
$\hat\Omega\ff$ is not  graded commutative this
element does not vanish!
\bb   h^{-1}\de h+h\de h^{-1}
\ne h^{-1}\de h+(\de h^{-1})h =  \de (h^{-1}h) = \de 1 =
0.\ee
Its image under $\pi_D$ however does vanish
\bb   \pi_D(h^{-1}\de h+h\de h^{-1}) &=&
\gamma(h^{-1}\de h+h\de h^{-1})\cr &=&
\gamma(h^{-1}\de h+(\de h^{-1})h) = 0.\ee
Therefore the considered element is in
$(\ker\pi_D)^1$ and the  corresponding element in
$\pi_D(\de(\ker\pi_D)^1)$ is
 \bb \pi_D(\de
h^{-1}\de h+\de h\de h^{-1}) &=&  \gamma(\de
h^{-1})\gamma(\de h) +\gamma(\de h)\gamma(\de
h^{-1})\cr
 &=& -\gamma^\mu\left({\partial\over{\partial
x^\mu}}h^{-1}\right)
     \gamma^\nu{\partial\over{\partial x^\nu}}h
  +\gamma^\nu\left({\partial\over{\partial x^\nu}}h
\right)
    \gamma^\mu{\partial\over{\partial
x^\mu}}h^{-1}\cr
 &=&
\lb\gamma^\mu\gamma^\nu+\gamma^\nu\gamma^
\mu\rb
       \left( {\partial\over{\partial x^\mu}}h^{-1}\right)
        {\partial\over{\partial x^\nu}}h\cr
 &=& -\left({2\over{h^2}}\eta^{\mu\nu}
           \left({\partial\over{\partial x^\mu}}h\right)
           {\partial\over{\partial x^\nu}}h\right) 1.\ee
 By
linear combination
\bb   \pi_D(\de(\ker\pi_D)^1) =
\left\{f1,\ \ f\in\ff \right\}.\ee
On the other hand
\bb   \pi_D(\hat\Omega^2\ff) =  \left\{
f_{\mu\nu}\gamma^\mu\gamma^\nu, \ \
f_{\mu\nu}\in\ff\right\}\ee
and
\bb   \pi_D(\de f_1\de
f_2+\de f_2\de f_1) =
\left(2\eta^{\mu\nu}{\partial\over{\partial
x^\mu}}f_1
           {\partial\over{\partial x^\nu}}f_2\right) 1\ee
After passage to the quotient $\pi_D(\de f_1)$ and
$\pi_D(\de f_2)$ anticommute whereas they did not
anticommute in $\pi_D(\hat\Omega^2\ff)$ and we
may now identify a general  element
\bb   \pi_D(f_0\de
f_1\de f_2)\ \ \in\Omega^2\ff\ee
 with the differential
2-form on $M$
\bb   f_0\de f_1\de f_2\ \ \in\Omega^2M.\ee
As in chapter 1 we have treated the quotient space like
a subspace  which is legitimate only in presence of an
appropriate scalar  product. Again this scalar product
will be defined in terms of the  involution and a trace.

The involution that $\Omega M$
inherits from $\Omega\ff$ via the  sketched
isomorphism is with our conventions
\bb   (f_0\de f_1\de
f_2...\de f_p)^* =  (-1)^{(1/2)p(p-1)}\bar f_0\de\bar
f_1\de\bar f_2...\de\bar f_p.\ee
The definition of a
trace is delicate because now our Hilbert space  $\sss$ is
infinite dimensional. For any bounded operator $Q$ on
$\sss$  we define the {\it Dixmier trace} $\tt$ by
\bb   \tt(Q|\ddd|^{-d}) := \lim_{N\rightarrow\infty}
                                {1\over{\log
N}}\sum_{n=1}^N\lambda_n\ee
 where $d=\dim M=4$
and the $\lambda_n$ are the eigenvalues of
 $Q|\ddd|^{-d}$
 arranged in a decreasing sequence discarding the zero
modes of the Dirac operator. Now we proceed
as in the finite  dimensional case ($d=0$) and define a
scalar product on  $\pi_D(\hat\Omega\ff)$ by
\bb   <\hat\phi_D,\hat\psi_D> := \tt
(\hat\phi^*_D\hat\psi_D|\ddd|^{-4}), \quad
\hat\phi_D, \hat\psi_D \in \pi_D(\hat\Omega\ff).\ee
Note that $\hat\phi_D$ is bounded because  $[\ddd,{\ul
f}]$ is. This scalar product can be computed to be
\bb   \label{one}
<\hat\phi_D,\hat\psi_D>= {1\over{32\pi^2}}\int_M\t_4
\lb\hat\phi_D^*\hat\psi_D\rb\de^4x\ee
independently of the four dimensional manifold $M$.
$\t_4$
denotes the trace over the gamma matrices.
 With this scalar product  $\Omega\ff$ is a subspace of
$\pi_D(\hat\Omega\ff)$ and by  definition orthogonal
to $J = \ker\pi_D + \de\ker\pi_D$. As subspace
$\Omega\ff$ inherits a scalar product $(\cdot,\cdot)$
given by
\bb   (\phi_D,\psi_D) = <\phi_D,P_D\psi_D>,
\quad \phi_D, \psi_D \in \Omega^p\ff\ee
where $P_D$
is the orthogonal projector in
$\pi_D(\hat\Omega\ff)$  onto the
ortho\--complement of $J$ and $\phi_D$ and $\psi_D$
are any  representatives in their classes. Thanks to
well known results for $\t_4\lb
\gamma^{\mu_1}v_{\mu_1}...
\gamma^{\mu_q}v_{\mu_q}\rb$
this scalar product now vanishes for forms with
different degree. By the above isomorphism between
$\Omega\ff$ and $\Omega M$ the differential forms
inherit a scalar  product still denoted by $ ( \cdot,\cdot
)$
 \bb   (\phi_D,\psi_D) = {1\over{8\pi^2}}
\int_M\phi_D^**\psi_D , \quad \phi_D, \psi_D \in
\Omega^pM\ee
 where the
Hodge star $*\cdot$ should not be confused with the
involution $\cdot^*$.

As an
example let us consider the flat 4-torus,
$M=T^4$ with all four circumferences measuring
$2\pi$ and let us compute the eigenvalues of the Dirac
operator
\bb \ddd\Psi = \pmatrix{
i{\partial\over{\partial x^0}}&0
&-{\partial\over{\partial x^3}}&
-{\partial\over{\partial x^1}}+
i{\partial\over{\partial x^2}}\cr
0&i{\partial\over{\partial x^0}}&
-{\partial\over{\partial x^1}}-
i{\partial\over{\partial x^2}}&
{\partial\over{\partial x^3}}\cr
{\partial\over{\partial x^3}}&
{\partial\over{\partial x^1}}-
i{\partial\over{\partial x^2}}&
-i{\partial\over{\partial x^0}}&0\cr
{\partial\over{\partial x^1}}+
i{\partial\over{\partial x^2}}&
-{\partial\over{\partial x^3}}&0&
-i{\partial\over{\partial x^0}} }
\pmatrix{
\Psi_1 \cr \Psi_2 \cr \Psi_3 \cr \Psi_4 }\ =\
\lambda\pmatrix{
\Psi_1 \cr \Psi_2 \cr \Psi_3 \cr \Psi_4 }.\ee
After a Fourier transform
\bb \Psi_A\ =:\ \sum_{j_0,...,j_3\in\zz}
c_A(j_0,...,j_3)\exp(-ij_\mu x^\mu),\quad A=1,2,3,4\ee
this equation reads
\bb \pmatrix{
j_0&0&ij_3&ij_1+j_2\cr
0&j_0&ij_1-j_2&-ij_3\cr
-ij_3&-ij_1-j_2&-j_0&0\cr
-ij_1+j_2&ij_3&0&-j_0}
\pmatrix{c_1\cr c_2\cr c_3\cr c_4}\ =\ \lambda
\pmatrix{c_1\cr c_2\cr c_3\cr c_4}. \ee
Its characteristic equation is
\bb \lb \lambda^2-(j_0^2+j_1^2+j_2^2+j_3^2)^2\rb^2\ =\
0\ee
and for fixed $j_\mu$ each eigenvalue
\bb \lambda\ =\ \pm\sqrt{j_0^2+j_1^2+j_2^2+j_3^2}  \ee
has multiplicity two. Therefore asymptoticly for large
$|\lambda|$ there are
$ 4|\lambda|^4B_4$ eigenvalues (counted with their
multiplicity) whose absolute values are smaller than
$|\lambda|$.
$ B_4=\pi^2/2$
denotes the volume of the unit ball
in $\rr^4$. Let us arrange the absolute values of the
eigenvalues in an increasing
sequence. Taking due account of their multiplicities
we have for large $n$
\bb |\lambda_n|\approx
\left({n\over{2\pi^2}}\right)^{1/4}\ee
and we can check the Dixmier trace in equation
(\ref{one}) for instance with $\hat\phi_D=\hat\psi_D
=1$
\bb <1,1>&=&\t_\omega(|\ddd|^{-4})=
\lim_{N\rightarrow\infty}
                                {1\over{\log
N}}\sum_{n=1}^N|\lambda_n|^{-4}\cr
&=&\lim_{N\rightarrow\infty}
                                {1\over{\log
N}}\sum_{n=1}^N{{2\pi^2}\over n}=
\lim_{N\rightarrow\infty}{1\over{\log
N}}\int_{1}^N{{2\pi^2}\over n}\,\de n\cr
&=&2\pi^2={1\over{32\pi^2}}\int_M\t_4[1]\de^4x.
\ee

Let us come back to the general case.
The group of unitaries is now infinite dimensional
\bb    G = \left\{g\in
\ff,\  gg^\ast =g^*g=1\right\} =
            \left\{M\rightarrow U(1)\right\},\ee
 the gauge
group $U(1)$. A gauge potential is simply a
differential  1-form $A$ with values in the Lie algebra
$u(1)$, the vector  potential  of electro\-magnetism.
Its curvature $F$ is the $u(1)$ valued 2-form
\bb   F :=
\de A+A^2 = \de A\ee
(the differential algebra
$\Omega\ff$ being graded  commutative $A^2\equiv
0$) and the Higgs potential is precisely the  Maxwell
action \bb   (F,F) = \int_MF*F.\ee

\section{ The tensor product}

 Remember the description of
spinning particles in  quantum mechanics. Particles
with spin $s$ come in $2s+1$ dimensional unitary
representations of the group $SU(2)$. Position in
space enters  the picture via the tensor product of this
finite dimensional Hilbert space with the space of
square integrable functions. In this spirit we shall
now turn the Higgses into genuine Higgs fields by
tensorizing $\aa$ and $\ff$.

 Let
us denote this tensor product by
\bb   \aa_t :=
\ff\otimes\aa.\ee
The algebra $\aa_t$ admits a natural
K-cycle $(\hh_t,\dd_t,\chi_t)$, the tensor product of
the  K-cycles $(\sss,\ddd,\gamma_5)$ on $\ff$ and
$(\hh,\dd,\chi)$ on  $\aa$. The Hilbert space
\bb   \hh_t
:= \sss\otimes\hh\ee
carries the representation
\bb   \rho_t := {\ul\cdot}\otimes\rho,\ee
 the chirality
operator is given by
\bb   \chi_t :=
\gamma_5\otimes\chi.\ee
The definition of the
generalized Dirac operator
\bb   \dd_t := \ddd\otimes 1 +
\gamma_5\otimes\dd\ee
 is well motivated from
differential geometry. We denote  by $\delta_t$ the
differential of the universal differential envelop
$\hat\Omega\aa_t$. This is only an auxiliary
construction and we shall never need $\delta_t$
expressed in terms of $\de$ and $\delta$. After passage
to the quotient the differential, still denoted by
$\delta_t$, will be the concrete operator
$-i\lb\dd_t,\cdot\rb$ which we shall have to calculate
in terms of $\ddd$ and $\dd$. To alleviate notations we
shall often omit the $\otimes$ and write e.g. $\dt fa$
for $\dt(f\otimes a)$. Again the good differential
algebra $\Omega\aa_t$ is obtained as a quotient via the
homomorphism
\bb   \pi_t(f_0a_0\dt f_1a_1...\dt f_pa_p)
:=
     (-i)^p\rho_t(f_0a_0)\lb\dd_t,\rho_t(f_1a_1)\rb...
                                    \lb\dd_t,\rho_t(f_pa_p)\rb.\ee
Our
basic variable $H_t$ is an antihermitian 1-form,
\bb   H_t
\in \Omega^1\aa_t,\qquad H_t^*=-H_t.\ee
Its curvature
is the hermitian 2-form
\bb   C_t:=\dt H_t + H_t^2\ee
used to
calculate the functional
\bb   V_t(H_t):=(C_t,C_t)\ee
 where
the scalar product now involves the Dixmier  trace in
$\sss$ and the trace in $\hh$. The main miracle of
Connes' recipe can be summarized in the
\hfil\break
{\bf Theorem:}
{\em
We have the folling decomposition
\bb
H_t=A+H,\qquad\qquad\quad\quad
A&\in&\Omega^1(M,\rho(\gg))
\hookrightarrow\Omega^1\ff\otimes\Omega^0\aa,\cr
                     H^*=-H&\in&\Omega^0(M,\Omega^1\aa)
\cong\Omega^0\ff\otimes\Omega^1\aa\ee
and
\bb C_t=F+C-D\Phi\gamma_5
\qquad\qquad\qquad\qquad\qquad\qquad
\qquad\qquad\ee
 with $\gg$
the Lie algebra of the group of unitaries
\bb   \gg:=\left\{X\in\aa,X^*=-X\right\},\ee
with the field
strength
\bb   F:=\de A+A^2=\de A+{1\over 2}[A,A]\ \
\in\Omega^2(M,\rho(\gg)), \ee
 and the covariant derivative
\bb   D\Phi:=\de\Phi+[A\Phi-\Phi A]\quad  \
\in\Omega^1(M,\Omega^1\aa).\ee
 Recall $\Phi := H-i\dd$. Finally the generalized Higgs
potential reads
 \bb   V_t(A+H)=\int_M\t(F*F)+\int_M\t(D\Phi^**D\Phi)
   +\int_M*V(H)-\int_M*V_0(H)\ee
 with
\bb   V_0(H):=<\alpha C,\alpha C>\ee
 and $\alpha$ is the
linear map
\bb   \alpha: \Omega^2\aa\longrightarrow
          \lb\rho(\aa)+\pi(\delta(\ker\pi)^1)\rb^\cc\ee
determined by the two equations
\bb
<R,C-\alpha C>&=&0\qquad{\rm for\ all}\
R\in\rho(\aa)^\cc, \\
 <K,\alpha C>&=&0\qquad {\rm for\ all}\
K\in\pi(\delta(\ker\pi)^1)^\cc.\ee
The scalar product
is the finite dimensional one of chapter 1, the $x$
dependence of  $C$ can be ignored.
    }

We have two basic variables. $A$ is
a genuine gauge  potential (non-abelian if $\aa$ is
non-commutative) that is a differential 1-form on $M$
with values in the Lie algebra $\gg$ of the group of
unitaries of $\aa$ represented on $\hh$. $H$ is as
before, however now with a differentiable $x$
dependence, i.e. $H$ is a multiplet of genuine scalar
fields. The generalized Higgs potential reproduces the
complete bosonic action of a Yang-\-Mills-\-Higgs
model, namely the Yang-\-Mills action, the covariant
Klein-\-Gordon action and the integral of the modified
Higgs potential $V-V_0$. The modified potential is still
a non-negative polynomial of fourth order in the
scalar fields, in fact, as we shall see
\bb   V-V_0=\t\lb(C-\alpha C)^2\rb.\ee
 The entire action is gauge invariant. An element $g$
of the group of unitaries of $\aa_t$ is a differentiable
function from spacetime into the finite dimensional
group $G= \{g\in \aa,\quad  gg^\ast
=g^*g=1\}$. Therefore these elements are genuine
gauge transformations. Under the gauge group our
fields transform as
\bb A^g &=
\rho(g)A\rho(g^{-1})+\rho(g)\de \rho(g^{-1}) ,\\
H^g &= \rho(g)H\rho(g^{-1})+\rho(g)\delta
\rho(g^{-1}).\ee
Before proving the theorem let us
compute the modified Higgs potential $V_0$ for our
example $\aa=M_2(\cc)\oplus\cc$. Recall the generic
elements
\bb   R=\pmatrix {a&0\cr 0&B}, \quad a \in
M_2(\cc),\ B=\pmatrix {b_1&0\cr 0&b_2}, \
b_1,b_2\in\cc, \ee
Note that the entries $b_1$ and $b_2$ of $B$ are now
unrelated due to the complexification
$\rho(\aa)^\cc$.
\bb   K=\pmatrix {\Delta k&0\cr
0&0}, \quad k \in M_2(\cc), \ee
 \bb   C=\pmatrix {0&0\cr
0&M^*cM}, \quad c \in M_2(\cc).\ee
Therefore
\bb   \alpha C=\pmatrix{ 0&0&0&0\cr 0&0&0&0\cr
0&0&|m_1|^2c_{11}&0\cr 0&0&0&|m_2|^2c_{22}},\ee
\bb V_0 &=&\ \t\lb(\alpha C)^2\rb\cr
&=&\ |m_1|^4(c_{11})^2+|m_2|^4(c_{22})^2\cr
        &=&\ |m_1|^4(1-\phi_1^*\phi_1)^2+
            |m_2|^4(1-\phi_2^*\phi_2)^2 \cr
        &=&\ |m_1|^4+|m_2|^4+|m_1|^4(\phi_1^*\phi_1)^2+
                                           |m_2|^4(\phi_2^*\phi_2)^2 \cr
          &   &\quad-2|m_1|^4\phi_1^*\phi_1
                -2|m_2|^4\phi_2^*\phi_2,\ee
 and the modified
Higgs potential reads
\bb   V-V_0=
2|m_1|^2|m_2|^2(\phi_1^*\phi_2)(\phi_2^*\phi_1).\ee
Through the process of tensorizing we have lost the
precious property of spontaneous symmetry breaking.
Indeed, the gauge invariant point $\Phi=0$ is minimum
of the modified potential. However,
this loss of symmetry breaking is particular to our
example and not a typical feature. In fact a slight
modification of the example namely adding more
families remedies the evil. Consider the same algebra
$\aa=M_2(\cc)\oplus\cc$ with a new Hilbert space
consisting of $N$ copies of the old one,
\bb   \hh:=(\cc^2\oplus\cc^2)\otimes\cc^N.\ee
E.g. for two
families, $N=2$, the representation is by the $8\times 8$
matrices
\bb   \rho(a,b):=\pmatrix{ a&0&0&0\cr
0&a&0&0\cr 0&0&B&0\cr 0&0&0&B}\ee
if written with respect to the suggestive basis
\bb (u_L,d_L,c_L,s_L,u_R,d_R,c_R,s_R) .\ee
The mass matrix
\bb   M=\pmatrix{ m_1&0\cr 0&m_2}\ee
has now entries
$m_j$, non-degenerate complex $N\times N$ matrices
which should be thought of as mass matrices of the
quarks of charge 2/3 and $-1/3$.
 In other words, the basis is ordered differently here:
\bb (u_L,c_L,d_L,s_L,\cdots),\quad N=2.\ee
The formulas of chapter one generalize naturally to
the new situation,
 \bb   MM^* = \pmatrix {m_1m_1^*&0\cr 0&m_2m_2^*} =
1\otimes\Sigma+\sigma_3\otimes\Delta\ee
 with
\bb   \Sigma := {1\over2}(m_1m_1^*+m_2m_2^*),\ee
\bb   \Delta :=   {1\over2}(m_1m_1^*-m_2m_2^*).\ee
The
generic elements $R\in\rho(\aa)^\cc, \
K\in\pi(\delta(\ker\pi)^1)$ and  $C\in \Omega^2\aa$
become
\bb   R=\pmatrix { a\otimes 1&0\cr
 0&B\otimes 1}, \quad a \in M_2(\cc),\ B=\pmatrix {
b_1&0\cr 0&b_2}, \ b_1,b_2\in\cc, \ee
\bb   K=\pmatrix
{k\otimes\Delta&0\cr 0&0}, \quad k \in M_2(\cc), \ee
\bb   C=\pmatrix {\tilde c\otimes\Sigma^\prime&0\cr
0&M^*(c\otimes 1)M}\ee
 with
\bb   c:=1-\phi^*\phi,\ee
\bb   \tilde c:=1-\phi\phi^*,\ee
\bb   \Sigma^\prime:=\Sigma-{{\t(\Delta\Sigma)}\over
{\t(\Delta^2)}}\Delta.\ee
A straight-forward calculation
yields
\bb   \alpha C=\pmatrix{ \tilde c\otimes Y&0\cr
0&z\otimes 1}\ee
with
\bb   Y:={{\t\Sigma'}\over
{N-((\t\Delta)^2/\t\Delta^2)}} \lb
1-{{\t\Delta}\over{\t\Delta^2}}\Delta\rb,\ee
\bb  z:={1\over N}\pmatrix { \t(m_1^*m_1)c_{11}&0\cr
0&\t(m_2^*m_2)c_{22}}\ee
and the (modified) Higgs
potential
\bb V-V_0&=&\t(\Sigma'-Y)^2\lb
(1-\phi_1^*\phi_1)^2+
                   (1-\phi_2^*\phi_2)^2+
                     2(\phi_1^*\phi_2)(\phi_2^*\phi_1)\rb
 \cr& &\quad+\lb\t(m_1^*m_1)^2-{1\over
N}(\t m_1^*m_1)^2\rb
         (1-\phi_1^*\phi_1)^2\cr &&\quad
+\lb\t(m_2^*m_2)^2-{1\over N}(\t m_2^*m_2)^2\rb
         (1-\phi_2^*\phi_2)^2  \cr &&\quad
-2\t(m_1^*m_1m_2^*m_2)
                 (\phi_1^*\phi_2)(\phi_2^*\phi_1)\ee
 does break the symmetry spontaneously unless there
is a numerical coincidence:
\bb   Y=\Sigma'\ \ {\rm and}\ \
z\otimes 1= \pmatrix { c_{11}m_1^*m_1&0\cr
0&c_{22}m_2^*m_2}.\ee

{\bf
Proof of the theorem:}
\hfil\break\noindent
 Our
conventions are summarized in table 1. In the
following we compute only $\Omega^1 \aa_t$ and
$\Omega^2 \aa_t$ for the particular case at hand.
Details of the general case can be found in reference
[2]. To get started,  we need the commutator
 $\lb\dd_t,{\ul f}\rho(a)\rb$. Using
\bb   \dd_t= \pmatrix
{ \ddd&\gamma_5M\cr
 \gamma_5M^*&\ddd}\ee
 we obtain
\bb   \lb\dd_t,{\ul
f}\rho(a)\rb=i\gamma(\de f)\rho(a)+
     \gamma_5{\ul f}\lb\dd,\rho(a)\rb\ee
and
\bb   \pi_t(f_0a_0\dt f_1a_1)=
   \pi_D(f_0\de f_1)\rho(a_0a_1)+
       \ul{f_0f_1}\gamma_5 \pi(a_0\delta a_1).\ee
Therefore $\Omega^1\aa_t$ decomposes into a direct
sum
\bb
\Omega^1\aa_t=\Omega^1\ff\otimes\rho(\aa)
\oplus\ff\otimes\Omega^1\aa\ee
 and its general
element can be put under the form
\bb   H_t=A+H,\ee
\bb   A=\pi_t(f_0a\dt f_11)\quad
\in\Omega^1\ff\otimes\rho(\aa)\ee
is a 1-form on
spacetime with values in $\rho(\aa)$ and
 \bb   H=\pi_t(fa_0\dt 1a_1)\quad
\in\ff\otimes\Omega^1\aa\ee
is an internal 1-form as
in chapter 1 but now with a differential $x$
dependence. Later we shall impose that $H_t$ be
antihermitian, in particular that $A$ take values in
the Lie algebra $\gg$
\bb   \gg:=\left\{X\in\aa,\quad X^*=-X\right\}\ee
represented on $\hh$ and $H$ is antihermitian.

On level two we have
\bb
\pi_t(f_0a_0\dt f_1a_1\dt f_2a_2)&=&
   \pi_D(f_0\de f_1\de f_2)\rho(a_0a_1a_2)
       +\ul{f_0f_1f_2}\pi(a_0\delta a_1\delta a_2)\cr
&&+i\gamma(f_0f_1\de f_2)\gamma_5\rho(a_0)
    \lb\dd,\rho(a_1)\rb\rho(a_2) \cr
&&-i\gamma(f_0\de
f_1f_2)\gamma_5\rho(a_0a_1)
    \lb\dd,\rho(a_2)\rb.\ee
Consequently
\bb
\pi_t(\hat\Omega^2\aa_t)=\lb\pi_D(\hat\Omega^2\ff)
    \otimes\rho(\aa)
       +\ff\otimes\pi(\hat\Omega^2\aa)\rb\oplus\
  \pi_D(\hat\Omega^1\ff)
\otimes\pi(\hat\Omega^1\aa).\ee
Remember from last
chapter that spacetime zero- and two-forms mix
before division by the junk. This entails that the sum
in the above bracket is not direct. Next we compute the
tensor junk. Consider a general element of
$\Omega^1\aa_t$
\bb   \pi_t(f_0a\dt f_11)+\pi_t(fa_0\dt 1a_1)\ee
 Its
pre-image $p$ belongs to $(\ker\pi_t)^1$ if and only
if $\pi_D(f_0\de f_1)=0$ and $\pi(a_0\delta a_1)=0$ in
which case
\bb   \pi_t(\delta_tp)=\pi_D(\de f_0\de
f_1)\rho(a)+ \ul f\pi(\delta a_0\delta a_1).\ee
 It follows that
\bb   \label{two}\pi_t(\dt(\ker\pi_t)^1)=
\pi_D(\de(\ker\pi_D)^1)\otimes\rho(\aa)
+\ff\otimes\pi(\delta(\ker\pi)^1)\ee
where again the
sum is not direct and by division we get
\bb
\Omega^2\aa_t=\lb\Omega^2\ff
    \otimes\rho(\aa)
       +\ff\otimes\Omega^2\aa\rb\oplus\
 \Omega^1\ff\otimes\Omega^1\aa.\ee
Our next task is to
compute the total curvature $C_t=\dt H_t+H_t^2$ with
$H_t=A+H$ and
\bb   A=\pi_t(f_0a\dt f_11)=\pi_D(f_0\de
f_1)\rho(a),\ee
\bb   H=\pi_t(fa_0\dt
1a_1)=\ul f\gamma_5\pi(a_0\delta
      a_1).\ee
We have
\bb   \dt A&=&\pi_t(\dt f_0a\dt f_11)=
\pi_D(\de f_0\de f_1)\rho(a)- \pi_D(f_0\de
f_1)\gamma_5\pi(\delta a)\cr& =:&\de A-\delta
A\gamma_5,\ee
\bb   \dt H&=&\pi_t(\dt fa_0\dt 1a_1)= \ul
f\pi(\delta a_0\delta a_1)+ \pi_D(\de
f)\gamma_5\pi(a_0\delta a_1)\cr&=:& \delta\tilde H+
\de\tilde H
\gamma_5,\ee
\bb (A+H)^2&=&\pi_D((f_0\de
f_1)^2)\rho(a^2) +\ul f^2\pi((a_0\delta a_1)^2)\cr
&&+\pi_D(f_0\de f_1f)\gamma_5\pi(aa_0\delta a_1)
  -\pi_D(ff_0\de f_1)\gamma_5\pi(a_0\delta a_1a)\cr
&=&:A^2+\tilde H^2+\lb A\tilde H-\tilde HA\rb
\gamma_5.\ee
 All together
\bb C_t&=&\de A-\delta
A\gamma_5+\delta\tilde H+\de\tilde H\gamma_5
+A^2+\tilde H^2+[A\tilde H-\tilde HA]\gamma_5\cr
&=&F+C+\left(i\lb\dd,A\rb+ \de\tilde
H+[A\tilde H-\tilde HA]\right)\gamma_5\cr
 &=&F+C-D\Phi\gamma_5\ee
with the curvatures
\bb   F:=\de A+A^2, \ee
\bb   C:=\delta\tilde H+\tilde H^2.\ee
 The covariant
derivative
\bb   D\Phi:=\de\Phi+[A\Phi-\Phi A]
\ \in\Omega\left(M,\Omega^1\aa\right)\ee
makes
sense because of the homogeneous transformation law
of $\Phi:=\tilde H-i\dd$. Note the purely algebraic term
$\delta A=-i[\dd,A]$ in the total curvature. This term
generates the masses of the gauge bosons in the
Lagrangian and does so by means of the fermionic
mass matrix $M$ in $\dd$.

Finally we have to work out the
scalar product in  $\Omega^2\aa_t$, the space where
the curvatures live. As  usual we start with the
auxiliary scalar product  $<\cdot,\cdot>$ in
$\pi_t(\hat\Omega\aa_t)$ defined by
\bb
<\hat\phi_D\otimes\hat\phi,\hat\psi_D\otimes
\hat\psi>&:=&
<\hat\phi_D,\hat\psi_D><\hat\phi,\hat\psi>=
\tt(\hat\phi_D^*\hat\psi_D|\ddd|^{-4})
\t(\hat\phi^*\hat\psi),\cr
&&\quad\hat\phi_D,\hat\psi_D\in\pi_D(\hat\Omega\ff),\
\hat\phi,\hat\psi\in\pi(\hat\Omega\aa).\ee
and
define the scalar product on $\Omega\aa_t$ by
\bb   (\phi_t,\psi_t):=<\phi_t,P_t\psi_t>,\ee
$\phi_t$ and
$\psi_t$ are any representatives of their classes and
$P_t$ is the orthogonal projector on $(\pi_t
(\dt(\ker\pi_t))^\perp$. For our purpose it is
sufficient to know this projector in
 $\pi_t(\hat\Omega^2\aa_t)$ where it is already non
trivial, $P_t\not=P_D\otimes P$. Let us come back to the
 decomposition
\bb
\pi_t(\hat\Omega^2\aa_t)=\lb\pi_D(\hat\Omega^2\ff)
    \otimes\rho(\aa)
       +\ff\otimes\pi(\hat\Omega^2\aa)\rb\oplus\
  \pi_D(\hat\Omega^1\ff)
\otimes\pi(\hat\Omega^1\aa)\ee
with general
elements in the three subspaces
\bb   \pi_t(f_0a_0\dt
f_11\dt f_21)= \pi_D(f_0\de f_1\de f_2)\rho(a_0),\ee
\bb   \pi_t(f_0a_0\dt 1a_1\dt 1a_2)= \ul f_0\pi(a_0\delta
a_1\delta a_2),\ee
 \bb   \pi_t(f_0a_0\dt 1a_1\dt f_21)=
-\pi_D(f_0\de f_2)\gamma_5\pi(a_0\delta a_1).\ee
 Our
first conclusion is that the above direct sum is also an
orthogonal sum because the trace in equation
(\ref{one}) is over
$\gamma_5$ multiplied by an odd number of proper
gamma matrices and vanishes. Therefore by
equation (\ref{two})
$P_t$ leaves the third subspace untouched and we
concentrate on the restriction $P_1$ of $P_t$ to the
first two subspaces. Let us introduce the following
short hands:
\bb   U:=\pi_D(\hat\Omega^2\ff)\otimes\rho(\aa),\ee
\bb   U_0:=\pi_D(\de(\ker\pi_D)^1)\otimes\rho(\aa)
\cong\ff\otimes\rho(\aa)\ee
and $U_\perp$ is the
ortho-\-complement of $U_0$ in $U$,
\bb   U_\perp:=(U_0)^{\perp
U}\cong\Omega^2M\otimes \rho(\aa).\ee
 Likewise
\bb   W:=\ff\otimes\pi(\hat\Omega^2\aa),\ee
\bb   W_0:=\ff\otimes\pi(\delta(\ker\pi)^1)\ee
and
$W_\perp$ is the ortho-\-complement of $W_0$ in $W$,
\bb   W_\perp:=(W_\perp)^{\perp W}=\ff\otimes\Omega^2
\aa\ee
and $P_1$ is the orthogonal projector in $U+W$
on $(U_0+W_0)^\perp\not= U_0^\perp+W_0^\perp$.
Next we remark that $U_\perp$ is orthogonal to the
other three subspaces
\bb   U_\perp\cap\left(U_0+(W_0\oplus
W_\perp)\right) =0,\ee
 \bb   U_\perp\perp\left(U_0+(W_0\oplus
W_\perp)\right) \ee
because
$\t_4[\gamma^\mu\gamma^\nu]=
\t_4[\gamma^\nu\gamma^\mu]$.
 This means that also $V_\perp$ decouples and it
remains to compute the restriction $P_2$ of $P_t$ to
\bb   U_0+(W_0\oplus W_\perp)=
\ff\otimes\lb\rho(\aa)+(\pi(\delta(\ker\pi)^1)
\oplus\Omega^2\aa)\rb.\ee
In this space now the $x$
dependence is trivial and can be ignored reducing the
calculation to a finite dimensional one in
$\lb\rho(\aa)+(\pi(\delta(\ker\pi)^1)
\oplus\Omega^2\aa)\rb $  with scalar product defined
by $\t$, the trace over $\hh$. What we need is $P_2C,\
C\in\Omega^2\aa$. Since $P_2$ is an orthogonal
projector onto $(U_0+W_0)^\perp$, $C-P_2C$ is
perpendicular to  $(U_0+W_0)^\perp$,
\bb   C-P_2C=:\alpha C\
\in(U_0+W_0)^{\perp\perp}=(U_0+W_0)^\cc\ee
 or
\bb   P_2C=C-\alpha C.\ee
 Finally by Pythagoras
\bb
<C,P_tC>&=&\cr
& &<C,P_2C>=<C,C-\alpha C>=(C,C)-<C,\alpha C>\cr
&&=(C,C)-<P_2C+\alpha C,\alpha C>\cr
&&= (C,C)-<\alpha
C,\alpha C>.\ee

\section{Remarks}

Instead of conclusions we offer a
list of remarks.

{\bf -
The standard model:\ }
 The first question is of course: can the standard model
be accommodated in Connes' approach? The answer is
yes, but this requires three new items to be added [1,3,4]
to Connes' model building kit as reviewed so far.
In our
example the group of unitaries,  $U(2)\times U(1)$, is
too big by one $U(1)$  factor to describe electroweak
interactions and there are two complex Higgs doublets,
one too many. Both points are cured readily by
replacing the internal algebra by $\aa=\hhh\oplus
\cc$, $\hhh$ being the quaternions, keeping the
K-cycle unchanged. Straightening out the
hypercharges requires a first new item, the so called
unimodularity condition which reduces the group of
unitaries by a purely algebraic restriction. But then
one has to start from a bigger algebra which is
anyhow needed to include strong interactions. Since
quantum chromodynamics is a vector like theory its
inclusion is easy except for two points. First its group,
$SU(3)$, is the group of unitaries of no algebra $\aa$.
This problem is taken care of by the unimodularity
condition. Second the quarks come in a
representation, that is a tensor product, $(3,2)$, and
therefore not available so far, a problem solved by
introducing bimodules, the second item. As explained
in our example the Higgs and gauge boson masses are
determined by the fermion mass matrix. Also it is
obvious that the coupling constants in the bosonic
action $V_t$ are related. In the complete standard
model \`a la Connes these relation are [5]
\bb   {g_3}={1\over2}\left({{4-2x}\over{1-x}}
\right)^{1/2}g_2\ee
 \bb   \sin^2\theta_W={{3(1-x/2)}\over{8-2x}}\ee
 and,
neglecting all fermion masses against the top mass
$m_t$:
\bb   m_W={1\over2}
\left({{(1-x)}\over{(1-x/2)}}\right)^{1/2}m_t\ee
 and
\bb   m_H=\left(3-{3\over2}{{3x^2-8x+5}
\over{5x^2-17x+14}}\right)^{1/2}m_t.\ee
 The
parameter $x$ ranges from$-1$ to +1. It is the relative
weight of the traces in the Hilbert spaces $\hh$ of the
leptons and of the quarks. Numerically \vskip.5cm

\bb   \vbox{\halign{\hfill#\hfill&\quad\hfill#\hfill&\quad\hfill#\hfill&\quad
\hfill#\hfill&\quad\hfill#\hfill&\quad\hfill#\cr
$x$&$-1$&0&${1\over2}$&0.99&1\cr
\noalign{\medskip} $\left({{g_3}/
g_2}\right)^2$&${3\over4}$&1&${3\over2}$&50.5&$\infty$\cr
\noalign{\medskip}
$\sin^2\theta_W$&${9\over{20}}$&${3\over8}$&${9\over{28}}$&0.2252&${1
\over4}$\cr \noalign{\medskip} ${m_t/
m_W}$&$\sqrt 3$&2&$\sqrt 6$&14.2&$\infty$\cr
\noalign{\medskip}
${m_H/m_W}$&$2.65$&$3.14$&$3.96$&$24.5$&$\infty$\cr}}\ee

Note that the ratio $m_H/m_t$
shows  little variation from  1.53 to $\sqrt 3$. Of course
these classical relations are unstable under quantum
corrections [6] and there are at least two possible attitudes
with respect to this dilemma. The first says we do not
know yet what quantum field theory is in this new
context. The other attitude is the third item: The trace
in the Hilbert space $\hh$ that we have used to define
the scalar product is not unique (up to normalisation)
and taking the most general traces produces the
standard model without the constraints on masses and
coupling constants.

{\bf - Other
generalisations:\ }
We have only considered a
particular case of Connes'
 algorithm. In general the total algebra $\aa_t$ is not a
tensor product of $\ff$ and a finite dimensional
algebra $\aa$. Also the basic variable $H_t$ can live in
a more general space than $\Omega^1\aa_t$, it can be a
connection on any hermitian finite projective module
over $\aa_t$. These generalisations do not seem to
help with the above mentioned problems.

{\bf - Other algebras:\ }
Up to now only few algebras $\aa$ have been explored
besides the standard model. It is most intriguing that
the simplest typical example we know so far is already
quite involved and reproduces most features of
electroweak interactions. The other algebras
considered in this context are $M_5(\cc)$ [7] and
 $Cliff(10)$ [8] in order to reproduce the $SU(5)$ and
$SO(10)$ grand unified theories. The first one fails
because the fermions in the $SU(5)$ model sit in a
$\bar 5+10$, the $10$ comes from a tensor product and
does not fit Connes' rules. The second model is left-\-
right symmetric and it is difficult to obtain the
complicated Higgs sector necessary in $SO(10)$.
Finally a smaller left-\-right symmetric model with
gauge group $U(2)\times U(2)$ has been worked out
[9]. As in the $SO(10)$ model, the gauge symmetry is
broken spontaneously and parity remains unbroken.

{\bf  - Splinters:\ } It
does not look easy to get rid of imaginary time. In non
compact, pseudo Riemannian spacetime traces and
scalar products are ill defined. The Dirac operator has
continuous spectrum, action integrals diverge and
there is certainly more to do than simply invoking
Wick rotation. There is no convincing motivation for
the time integral in $\sss=L^2(M,\cc^4)$.

{\bf - Other non
commutative schemes:\ }
Connes' algorithm to
produce a differential algebra  $\Omega\aa$ starting
from an algebra $\aa$ --- such that the algebra $\ff$
of functions on a manifold $M$ reproduces de Rham's
differential algebra $\Omega M$ --- is not unique. In
fact already in 1988  Dubois-\-Violette [10] introduced a
different such algorithm using derivations. This
algorithm is successfully used by the southern Paris
group to obtain particle models [11]. These models
share the attractive features of Connes' theory, they
also unify gauge and Higgs bosons. In their scheme
however the Higgs bosons transforms according to the
adjoint representation of the group of unitaries
irrespective of fermion representations. Their scheme
has been  generalized by Balakrishna,  G\"ursey \&
 Wali [12] to include other Higgs representations.
Another approach due to
Coquereaux [13] takes immediately the differential
algebra $\Omega\aa_t$ as starting point. The approach
is thereby more transparent and less rigid. Many of its
physical features have been worked out by the
Marseille-\-Mainz group [14].

{\bf - Non-commutative
algebras and quantum physics:\ } The
non-\-commutative models are clearly inspired by the
mathematics of quantum mechanics, operator algebras.
There are two main differences between non
commutative and quantum. Ad one, the passage from
classical to quantum mechanics can be considered as
replacing the commutative algebra of functions on
phase space, the ''observables'', by a non-commutative
algebra [15]. In the above models not phase space but
space time is rendered non-commutative. Ad two, in
quantum mechanics the non-commutative algebra is
God given and contains a dimensionful parameter,
Planck's constant. A nice example illustrating the
interplay of quantum mechanics and non-commutative
geometry is given by Madore [16]. We repeat that a
generalisation of quantum field theory to the
non-\-commutative setting is still lacking.

{\bf   - Parallel
universes:\ }   The first example studied by Connes and
Lott [1] was $\aa=\cc+\cc$, $\hh=\cc+\cc$. It has a nice
geometric interpretation in terms of a Riemannian
manifold $M=M_L+M_R$, disjoint union of two
identical Riemannian manifolds $M_L$ and $M_R$
separated by the constant distance
$(|m_1|^2+|m_2|^2)^{-1/2}$. The left handed fermions
live on $M_L$, the right handed fermions on $M_R$.
This model can also be interpreted as a Kaluza-\-Klein
theory with a discrete fifth dimension consisting of
two points. The Kaluza-\-Klein analogy is present in
all non-commutative models. It has been worked out in
detail by the southern Paris group [17] and served as
initial motivation for Coquereaux's scheme [13].

{\bf  - Gravity:\ }  A
question of fundamental importance is: can the
Einstein-\-Hilbert action be fit into the non
commutative frame? Again the answer is not unique.
A first proposal is due to the Z\"urich group [18].
Starting from Einstein-\-Cartan's theory they arrive at
a tensor scalar theory. The scalar has a geometric
interpretation as the now variable distance between
parallel universes. In a recent paper Chamseddine and
Fr\"ohlich [19] have coupled this scalar to the standard
model and after addition of some effective
Coleman-\-Weinberg potential they obtain a striking
prediction for the top and Higgs masses,
\bb   146.2\le m_t\le 147.4\ {\rm GeV},\ee
 \bb   117.3\le m_H\le 142.6\ {\rm GeV}.\ee
Connes [20] has found a more intrinsic way to
incorporate gravity: he computes the Dixmier trace or
more properly the Wodizicki residue of the (true) Dirac
operator to the power minus two and obtains the
Einstein-\-Hilbert action.

{\bf  - Unification:\ }
Grand unification was based on the attractive idea to
replace a direct product of groups by a simple group.
In the same spirit and in order to reconcile particle
interactions and gravity it seems attractive to look for
an algebra $\aa_t$, that is not just a tensor product of
the algebra $\ff$ of functions on spacetime and an
internal algebra, but that is still sufficiently close to
$\ff$ to allow spacetime to subsist in some form. A
2-dimensional example of such an algebra is the
non-\-commutative torus which plays an intriguing
role in solid state physics.

We are indebted to Daniel Kastler who introduced us to
the miraculous non-\-commutative world, he is a most
pleasant guide. It is also a pleasure to acknowledge
stimulating discussions with Robert Coquereaux, Gilles
Esposito-\-Far\`ese, Bruno Iochum and Daniel $\rm
T^*$.
 \vfil\eject

\centerline{\bf References}

\begin{description}
\item{\ [1]} A. Connes, {\it Non-Commutative
Geometry},
Publ. Math. IHES 62 (1985),\hfil\break
A. Connes \& J. Lott, Nucl. Phys. Proc. Suppl.
B18 (1989) 29, \hfil\break
 A. Connes, {\it Non-Commutative Geometry}, Academic
Press (1993)\\
 A. Connes \& J. Lott, {\it The metric
aspect of non-commutative geometry}, in the
proceedings of the 1991 Carg\`ese Summer Conference,
eds.: J. Fr\"ohlich et al., Plenum Press (1992)
\item{\ [2]} D. Kastler \& D. Testard, {\it Quatum
forms of tensor products}, Comm. Math. Phys. in press
\item{\ [3]} D. Kastler, A detailed account of Alain
 Connes' version of the standard model in
non-commutative differential geometry I, II, and III.
to appear in Rev.Math.Phys. \hfil\break
D. Kastler \& M. Mebkhout, {\it Lectures on
Non-Commutative Differential Geometry}, World
Scientific, to be published
\item{\ [4]} J.C. V\'arilly \&  J. M. Gracia-Bond\'\i a, {\it
Connes' noncommutative differential geometry and
the standard model}, J. Geom. Phys., in press
\item{\ [5]} D. Kastler \& T. Sch\"ucker,
Theor. Math. Phys. 92 (1992) 522
\item{\ [6]} J. Kubo, K. Sibold \& W. Zimmermann,
Nucl. Phys. B259 (1985) 331 \hfil\break
E. Alvarez, J. M. Gracia-Bond\'\i a \& C. P. Mart\'\i n,
Phys. Lett. 306B (1993) 55
\item{\ [7]} A. Chamseddine, G. Felder \& J. Fr\"ohlich,
Phys. Lett. 296B (1993) 109, Nucl. Phys. B395 (1993) 672
\item{\ [8]} A. Chamseddine \& J. Fr\"ohlich, {\it
$SO(10)$ Unification in non-commutative geometry},
ZU-TH-10/1993, ETH/TH/93-12
\item{\ [9]} B. Iochum \& T. Sch\"ucker, {\it A
left-right symmetric model \`a la Connes},
CPT-93/P.2973
 \item{[10]} M. Dubois-Violette, C. R. Acad. Sc.
Paris 307 I (1988) 403
\item{[11]} M. Dubois-Violette, R. Kerner \& J.
Madore, Phys. Lett. 217B (1989), Class. Quant. Grav. 6
(1989) 1709, J. Math. Phys 31 (1990) 316, J. Math. Phys.
31 (1990) 323, Class. Quant. Grav. 8 (1991) 1077,
\hfil\break H. Grosse \& J. Madore, Phys. Lett. 283B
(1992) 218, \hfil\break
J. Madore, Mod. Phys. Lett. A4 (1989) 2617,
J. Math. Phys. 32 (1991) 332,
Int. J. Mod. Phys. A6 (1991) 1287,
Phys. Lett. 305B (1993) 84,
{\it On a non-commutative
extension of Electro\-dynamics}, LPTHE Orsay 92/21
\item{[12]} B. S. Balakrishna, F. G\"ursey \&
K. C. Wali, Phys. Lett. 254B (1991) 430, Phys. Rev. D 44
(1991) 3313
\item{[13]} R. Coquereaux, G. Esposito-Far\`ese \&
G. Vaillant, Nucl. Phys.B353 (1991) 689
\item{[14]} R. Coquereaux, G. Esposito-Far\`ese \&
F. Scheck, Int. J. Mod. Phys. A7 (1992) 6555,
\hfil\break
R. H\"au\ss ling, N.A. Papadopoulos \& F. Scheck,
Phys. Lett. 260B (1991) 125,
Phys. Lett. 303B (1993) 265 \hfil\break
R. Coquereaux, R. H\"au\ss ling, N.A. Papadopoulos
\& F. Scheck, Int. J. Mod. Phys. A7 (1992) 2809
\hfil\break
F. Scheck, Phys. Lett. 284B (1992) 303,
\hfil\break
R. Coquereaux, R. H\"au\ss ling \& F. Scheck, {\it Algebraic
Connections on Parallel Universes}, preprint
Universities of Macquarie and Mainz 1993
\hfil\break
N.A. Papadopoulos, J. Plass \& F. Scheck, {\it Models of
elctroweak interactions in non-\-commutative
geometry: a comparison},
 MZ-TH/93-26
\item{[15]} M. Dubois-Violette in Differential
Geometric Methods in Theoretical Physics. Eds. C.
Bartocci et al. Springer, Berlin, 1991
 \item{[16]} J.
Madore, Phys. Lett. 263B (1991) 245, Class. Quant. Grav. 9
(1992) 69, Ann. Phys. 219 (1992) 187
\item{[17]} M.
Dubois-Violette, R. Kerner \& J. Madore,
 Class. Quant. Grav. 6 (1989) 1709, \hfil\break
J. Madore, Phys. Rev. D 41 (1990) 3709, \hfil\break
J. Madore \& J. Mourad, {\it Algebraic Kaluza-\-Klein
cosmology}, LPTHE Orsay 93/000
\item{[18]} A. Chamseddine, G. Felder \& J. Fr\"ohlich,
{\it Gravity in non-commutative geometry}, Comm. Math.
Phys. to appear
\item{[19]} A. Chamseddine \& J. Fr\"ohlich,
{\it Constraints on the Higgs and top quark masses from
effective potential and non-commutative geometry},
ZU-TH-16/1993
\item{[20]} A. Connes to be published in the
proceedings of the 1992 Les Houches Summer School
\end{description}
\vfil\eject
\end{document}